\documentstyle[12pt,fleqn,cite]{article}

\def\be{\begin{equation}}
\def\ee{\end{equation}}
\def\bea{\begin{eqnarray}}
\def\eea{\end{eqnarray}}

\begin{document}
\begin{titlepage}
\begin{center}
\hfill hep-th/9801141\\
\hfill IP/BBSR/98-03\\
\vskip .2in

{\Large \bf String Network and U-Duality}
\vskip .5in

{\bf Sandip Bhattacharyya, Alok Kumar and Subir Mukhopadhyay\\
\vskip .1in
{\em Institute of Physics,\\
Bhubaneswar 751 005, INDIA}}
\end{center}

\begin{center} {\bf ABSTRACT}
\end{center}
\begin{quotation}\noindent
\baselineskip 10pt
We discuss the generalization of recently discovered 
BPS configurations, corresponding to the planar 
string networks, to non-planar
ones by considering the U-duality symmetry of 
type II string theory in various dimensions.  
As an explicit example, we analyze the string solutions in 
8-dimensional space-time, carrying SL(3) charges, and show that
by aligning the strings along various directions appropriately, one can 
obtain a string network which preserves 1/8 supersymmetry. 
\end{quotation}
\vskip .2in
IP/BBSR/98-03\\
January 1998\\
\end{titlepage}
\vfill
\eject


BPS solutions of string theory have played a crucial role in 
understanding its strong coupling aspects as well as in 
formulating the black hole information loss paradox in 
string theory context. In this context, a new class of
BPS configuration has been discovered recently. They have
the structure of a string network\cite{sennet}. The 
occurance of similar networks also takes place in other theories
of gravity\cite{rovel,smo2,smolin}. 
A possible role of string networks in compactifications, which 
take into account the non-perturbative aspects, has been 
pointed out \cite{sennet} and their dynamics has been 
discussed in \cite{rey,smolin}. 
In view of their varied applications, it is 
of interest to investigate the possibility of more 
general string networks. In this context, 
in this note, we show the existence
of non-planar networks in string theory.

The basic building blocks for these networks are the 3-string 
junctions\cite{schwarz,ahaha,aha,zweib,dasg,berg,lee,mats} 
of type IIB strings carrying $B_{\mu \nu}$ charges 
(p, q)\cite{sl2}, where $(p, q)$ denote the charges 
with respect to the antisymmetric tensor fields,
$B_{\mu \nu}^i$ (i=1,2), from the 
NS-NS and R-R sectors respectively. These junctions act as the
vertices in the network. On each of the vertices,
the charges are restricted by the condition:
$\sum_i p_i = \sum_i q_i = 0$\cite{schwarz,dasg,sennet}. 
The BPS nature of these junctions 
was conjectured earlier. However, a 
more explicit argument of its stability, based on supersymmetry,
has been presented only recently\cite{dasg,sennet}. 
Based on an earlier conjecture\cite{schwarz}, 
string junctions and networks have also been 
analyzed from the 11-dimensional M-Theory point of view\cite{lee}, 
where they can be embedded into a membrane solution.

In this paper, we consider the possibility of non-planar 
string networks of strings which carry $B_{\mu \nu}$ 
charges with respect to several (more than two)
antisymmetric tensor 
fields. These are not possible in ten dimensions. However,
as one compactifies the type IIA or IIB string theories further,
one gets several such degrees of freedom. They transform 
as a representation of the corresponding U-duality 
group. For example, in 8-dimensions, the full U-duality 
group is given as $SL(3, Z)\times SL(2, Z)$ and three 
antisymmetric tensor fields transform as a $(3, 1)$ representation
of this symmetry. Similar results hold in all lower dimensions
as well. In this paper, for explicit illustrations we will 
restrict ourselves to the eight dimensional case.

It is also known that the string solutions which carry 
$B_{\mu \nu}$ charges exist in all space-time dimensions 
$D\geq 4$. These are classified by the solution
of the Green's function equation for the corresponding
transverse space. The Green's function are 
determined in terms of the string source. For the ten 
dimensional type IIB string theory, it has been shown 
explicitly that, given a $(1, 0)$ type string, one 
can obtain a $(p, q)$ type solution through a series of 
$SL(2)$ transformations\cite{sl2}. The string tension for a 
$(p, q)$ string has an $SL(2)$ invariant form, 
which, in addition, is specified by the asymptotic 
value of the moduli. 

Based on the structure of the string solution, one can argue that 
a similar exercise is possible in lower dimensions as well 
and it is possible to generate string solutions 
in 8-dimensions which carry the charges with respect to 
all the three antisymmetric tensors. These string solutions
can then be classified by three integers $(p_1, p_2, p_3)$. 
The string tension now has an $SL(3)$ invariant form
and are functions of these charges as well as the 
$SL(3)/SO(3)$ moduli. Since $B_{\mu \nu}$'s 
are singlets with respect to the
$SL(2)$ part of the U-duality, this tension is also an 
$SL(2)$ singlet. Brane configurations, 
carrying U-duality charges, and their geometric interpretation 
has also been discussed in \cite{leung}.

Now, to construct a stable network of
strings carrying such charges, we analyze the supersymmetry 
conditions and show that a certain amount of supersymmetry is
indeed preserved in the present case. This is true provided 
the orientation of various strings in a three dimensional 
space, specified below by three directions, 
$x^5..x^7$, is coupled to its orientation in a three
dimensional internal space, specified by a ``charge vector''.
This charge vector, in turn, is specified by  
the values of the $SL(3)$ moduli and 
the $SL(3)$ charges in an appropriate way. 
Similar alignment of string in internal and space-time directions
was also necessary for constructing 
the planar string network discussed in 
\cite{sennet} and reminds of the $F$-Theory 
constructions\cite{vafa},
where the space-time direction plays the role of the base of a 
manifold and the U-duality directions play the role of its
fiber. 

As is known, the supercharges transform only with respect to the 
maximal compact subgroup of the U-duality group. In the
present case they form the representation: 
$Q_{1/2} \sim (S_8^+, 2)_1 + (S_8^-, 2)_{-1}$, with 
$S_8^+ (S_8^-)$ denoting the spinor representations of the
Lorentz group and other indices denote the transformation
property with respect to the maximal compact 
subgroup, $SU(2) \times U(1)$, of the U-duality group. 
$SU(2) \equiv SO(3)$ is the maximal compact subgroup of 
$SL(3)$ and $U(1)$ is the maximal compact subgroup of 
$SL(2)$. We represent the corresponding supersymmetry 
transformation parameters by spinors: $\epsilon_+^i$ and 
$\epsilon_-^i$, with (i =1, 2), respectively. 
For simplicity, we only 
analyze the sector of the theory containing the 
graviton, 3-antisymmetric tensors and $SL(3)$ moduli. 
In other words, all other moduli are set to zero and 
the states are neutral with respect to all other charges
like third-rank antisymmetric tensor and gauge fields.

We now follow the line of argument in \cite{sennet} and  
write down a U-duality invariant 
supersymmetry condition, preserved by a string carrying
charges $(p_1, p_2, p_3)$, in the present case as:
\bea
	\pmatrix{\epsilon_{\pm}^1 \cr
	\epsilon_{\pm}^2} = -i {{\cal X} \over |X|}
	\Gamma_1 ... \Gamma_6
	\pmatrix{\epsilon_{\pm}^1 \cr
		 \epsilon_{\pm}^2},
			\label{onehalf}
\eea
where, as mentioned earlier, $\epsilon_{\pm}^i$ are the 
positive (negative) 
chirality spinors transforming as a doublet of 
$SU(2)$, the maximal compact subgroup of $SL(3)$. 
$\Gamma_i$ are the eight dimensional gamma matrices and 
${\cal X} \equiv X_i \sigma_i$ is a $2\times 2$ matrix
defined in terms of a ``charge vector'' with components $X_i$'s
(i=1..3). $|X|$ is the magnitude of this vector. The factor 
$i$ is needed, since $(\Gamma_1...\Gamma_6)^2 = -1$.

We now construct an explicit form for the three components $X_i$'s,
of the ``charge vector'' $\vec{X}$ in terms of the $SL(3)/SO(3)$
moduli $G$ and the integers $(p_1, p_2, p_3)$ denoting the 
$B_{\mu \nu}$ charges. These moduli fields in the present case
can be paramaterized by a real symmetric matrix with unit 
determinant:
\bea
	G = \pmatrix{ g + a^T a e^{-\phi} & 
		      e^{-\phi}a^T \cr
		a e^{-\phi} & e^{-\phi} },
			\label{moduli}
\eea
where $g$ is a $2\times 2$ matrix:
\bea 
	g = \pmatrix{ e^{(\phi + \alpha)} 
		+ \chi^2 e^{-\alpha} & 
		e^{-\alpha} \chi \cr
		\chi e^{-\alpha} & e^{-\alpha}}
		\label{gnot}
\eea
and $a$ is a 2-dimensional row vector with components
$(a_1, a_2)$. It can be 
verified that the matrix $G$, with $g$ defined in 
eqn. (\ref{gnot}) is a (real, symmetric)
matrix with unit determinant. It is 
parameterized by five real parameters $(a_1, a_2)$, $\chi$,
$\phi$ and $\alpha$. In writing down the string tension and 
the charge vector, we use the asymptotic values of these 
moduli, denoted below by their subscripts $0$. 
The components of the charge vector are then written as: 
\be 
	X_i = ({\lambda_0}^{-1})_{i a} p_a
				\label{charge}
\ee
where $\lambda^{-1}$ are the ``vielbein's'' satisfying 
$\lambda \lambda^T = G$ and the index $(a=1..3)$. 
Explicitly, we have:
\bea
	\lambda^{-1} = \pmatrix{ e^{-{(\phi + \alpha)/2}}
	& - \chi e^{- {(\phi + \alpha)/2}} &
	- e^{- {(\phi + \alpha)/2}} a_1 + 
	\chi e^{- {(\phi + \alpha)/2}} a_2	\cr
	0 & e^{\alpha/2} & - e^{\alpha/2}a_2 \cr
	0 & 0 & e^{\phi/2} }.
		\label{lambda}
\eea
One obtains two $SL(2)$ subgroups by imposing the conditions:
(i) $\phi$ = $a_1$ = $a_2 = 0 $ and (ii) $\alpha = -\phi$, 
$\chi = a_1 = 0$. The form of $\lambda^{-1}$ for these two cases
then match with the expressions given in \cite{sl2}. 

Now, since the spinors transform 
in a fundamental representation: 
\be
   \pmatrix{\epsilon^1_{\pm} \cr
	    \epsilon^2_{\pm} }  \rightarrow 
	U \pmatrix{\epsilon^1_{\pm} \cr
	    \epsilon^2_{\pm} },
			\label{transform}
\ee
the supersymmetry condition in eqn.(\ref{onehalf})
is going to remain invariant provided
\be 
	U {\cal X}' U^{\dagger} = {\cal X}.
				\label{calx}
\ee
In other words, the transformation of the spinors are 
absorbed in that of the vector $X_i$'s. This is possible,
since for Pauli matrices appearing in 
the definition of ${\cal X}$, we have 
\be 
	U \sigma_i U^{\dagger} = O_{i j} \sigma_j,
		\label{sigma}
\ee
where $O$ is a 3-dimensional rotation matrix. To see this
in a more explicit manner, one can parameterize the $SU(2)$
matrix $U$ through Euler-angles:
\be
	U = exp.(i\phi_L \sigma_2) exp.(i \rho \sigma_3)
		exp.(i \phi_R \sigma_2)
		\label{euler}
\ee
and obtain the standard expression for the rotation matrix
$O$. The components of the charge vector 
transform as:
 \be
	X'_i = O_{i j} X_j
		\label{x'}
\ee
with $O$ being an $SO(3)$ rotation matrix. 
The transformation
(\ref{x'}) of the charge vector can be seen from its definition
(\ref{charge}) and by noticing that the vielbein $\lambda$ 
transforms as a vector, both under $SL(3)$ and $SO(3)$, namely 
the ``world''-indices and the ``tangent-space'' indices. 
The ``world'' or $SL(3)$ indices
are then contracted with the vector $\vec{p}$ denoting the 
$B_{\mu \nu}$ charges, and we are left with an $SO(3)$ vector. 
In other words, the proof of the invariance of the supersymmetry
condition (\ref{onehalf}) follows from the fact that the 
the unitary rotation of the
spinors are absorbed in the orthogonal rotations of the vectors, 
just as in the Dirac theory. 

After showing the U-duality invariance of the supersymmetry condition,
we now analyze the possibility of having several strings, 
aligned along various directions, in the eight dimensional space-time. 
In our case, we consider the possibility of
three dimensional spaces, spanned by the directions $x^i$, (i=5,6,7). 
For a string oriented along a direction making an angle $\theta$ with 
respect to the $x^7$ axis and angle $\phi$ with respect to the 
$x^6$ axis in $x^5$, $x^6$ plane, the supersymmetry
condition takes the form:
\bea
	\pmatrix{\epsilon_{\pm}^1 \cr
		 \epsilon_{\pm}^2} = -i {{\cal X}\over |X|} 
	\Gamma_1 ... \Gamma_4
	(\Gamma_5 \Gamma_6 cos \theta + \Gamma_5 \Gamma_7 sin\theta
	cos\phi + \Gamma_6 \Gamma_7 sin\theta sin\phi)
	\pmatrix{\epsilon_{\pm}^1 \cr
		 \epsilon_{\pm}^2},
			\label{super}
\eea
To solve this condition, we rewrite ${\cal X}$ as 
\be
	{\cal X} = |X|(\sigma_3 cos\theta + 
	\sigma_1 sin\theta cos\phi +
	\sigma_2 sin\theta sin \phi)
		\label{align}
\ee
The condition (\ref{align}) in fact gives the alignment of 
string in the internal space with respect to the one in 
the real space-time. The above supersymmetry 
condition can then be solved for arbitrary 
$\theta$ and $\phi$ by imposing the
following set of conditions:
\bea
	\Gamma_1...\Gamma_6\pmatrix{\epsilon^1_{\pm} \cr
	\epsilon^2_{\pm}}  = i \sigma_3 \pmatrix{\epsilon^1_{\pm} \cr
	\epsilon^2_{\pm}} ,	
\quad
	\Gamma_1...\Gamma_5\Gamma_7\pmatrix{\epsilon^1_{\pm} \cr
	\epsilon^2_{\pm}}  = i \sigma_1 \pmatrix{\epsilon^1_{\pm} \cr
	\epsilon^2_{\pm}} , 
\label{cond2} 
\eea	
\bea
	\Gamma_1...\Gamma_4\Gamma_6\Gamma_7
	\pmatrix{\epsilon^1_{\pm} \cr
	\epsilon^2_{\pm}}  = i \sigma_2 \pmatrix{\epsilon^1_{\pm} \cr
	\epsilon^2_{\pm}} 	\label{cond3}
\eea
One can now show in various ways that the 
conditions (\ref{cond2}),(\ref{cond3}) preserve
a part of the original supersymmetry.
First,  it can be shown that they are
equivalent to another set of conditions namely:
\bea \!\!\!\!\!\!\!\!
	\Gamma_1...\Gamma_4\pmatrix{\epsilon^1_{\pm} \cr
	\epsilon^2_{\pm}}  =  \pmatrix{\epsilon^1_{\pm} \cr
	\epsilon^2_{\pm}} 
\quad
	\Gamma_5\Gamma_6\pmatrix{\epsilon^1_{\pm} \cr
	\epsilon^2_{\pm}}  = i \sigma_3 
	\pmatrix{\epsilon^1_{\pm} \cr
	\epsilon^2_{\pm}} 
\quad
	\Gamma_5\Gamma_7\pmatrix{\epsilon^1_{\pm} \cr
	\epsilon^2_{\pm}}  = i \sigma_1 
	\pmatrix{\epsilon^1_{\pm} \cr
	\epsilon^2_{\pm}}. \label{cond3'}
\eea
Now, for the eight dimensional spinor of 
a given helicity, the first condition of (\ref{cond3'}) fixes the helicity 
in a four dimensional subspace. The second condition of (\ref{cond3'}) chooses 
one particular helicity in the remaining four dimensional space.
Therefore, for a spinor of given chirality in the 8-dimensional space-time,
this fixes the chiralities in the  
first as well as the last four dimensional subspaces. The last 
condition of (\ref{cond3'}) then keeps only a linear combination of these
two spinors, $\epsilon^1$ and $\epsilon^2$,  invariant. 
As a result, we have shown that by orienting 
strings along various directions in the 3-dimensional space, we
can obtain a solution which preserves $1/8$ supersymmetry.

We have also examined the supersymmetry condition more explicitly
by choosing the 8-dimensional $\Gamma$ matrices in Majorana 
representation\cite{gsw}. 
This combines $\epsilon_+^i$ and
$\epsilon_-^i$ into the spinors $\epsilon^i$ (i=1,2) 
as the two components of $\Gamma_9$ . 
The three relevant products of $\Gamma$ matrices, 
in this representation, are given by
\bea
\Gamma_1\cdots\Gamma_4\Gamma_5\Gamma_6 = - 
i\sigma_2\times 1\times\sigma_3\times 1 = l_3\times 1\nonumber, \\
\Gamma_1\cdots\Gamma_4\Gamma_6\Gamma_7 = -
\sigma_3\times i\sigma_2\times 1\times 1 = l_2\times 1\nonumber, \\
\Gamma_1\cdots\Gamma_4\Gamma_5\Gamma_7 = -
i\sigma_2\times 1\times\sigma_1\times 1 = l_1\times 1,
\eea
where we have introduced $8\times8$ matrices $l_i$ for convenience.
In terms of these matrices, the supersymmetry 
condition can be written as
\be \left(\begin{array}{cc}l_i & 0 \\ 0 & l_i \end{array}\right) 
\left( \begin{array}{c} \epsilon_+^1 \\ \epsilon_+^2\end{array}\right) = 
i \sigma_i\left( \begin{array}
{c}\epsilon_+^1 \\ \epsilon_+^2 \end{array} \right). \ee
The other chiral components $\epsilon_-^i$ satisfy similar relations.

A straightforward calculation then shows that these correspond to 
three independent relations given by
\be \epsilon_+^2 = -il_1\epsilon_+^1, \quad\quad 
l_1l_2l_3 \epsilon_+^1 = -\epsilon_+^1, \quad\quad
l_3 \epsilon_+^1 = i\epsilon_+^1 . \ee
Where $\epsilon_+^I$ are 8 component spinors. 
The first of these conditions then determines 
$\epsilon_+^2$ in terms of $\epsilon_+^1$ and 
thus breaks one half of the 
supersymmetry. Moreover, 
each of the remaining two conditions project out one 
half of the components of $\epsilon_+^1$. Thus we are 
left with $1/8$ supersymmetry.

We have therefore shown the supersymmetric nature of our solution. 
Now, since $1/8$ supersymmetry is preserved for arbitrary 
$\theta$ and $\phi$,
we conclude that several strings can be put together 
in a manner preserving the supersymmetry. The 
orientation of individual strings are dictated by the charge vectors,
which give the corresponding angles. These angles in turn are 
determined in terms of the $B_{\mu \nu}$ charges $p_i$ and 
the moduli $G_0$. Each 3-string junction,
acting as a vertex of the network,  still
forms a plane. However, the full network has a non-planar form,
as vectors meeting at different vertices can lie in different planes. 
The stability of the solution is also possible to show through 
the balance of forces at each of the vertices. 
This is because, like in the case of IIB strings in 10-dimensions,
the tension is given by an expression which is dependent only on the 
magnitude of the charge vector. The balance of forces then hold
through the usual vectorial sum.

To conclude, we have used the U-duality symmetry of 
the type II string theory in eight dimensions to construct
a generalized string network with non-planar structure. 
We have exploited the geometric nature of the $SL(3)$ 
U-duality symmetry for this purpose. It is possible to 
construct even more general string networks by using 
bigger U-duality symmetry in lower dimensions. However it
remains to be seen how much supersymmetry they preserve. 
One can possibly also analyze the supersymmetry properties,
when higher dimensional branes are involved in a similar 
manner. We hope to report on these issues in near future.


\vfil
\eject

\end{document}